\documentclass[10pt]{article}
\usepackage{amsfonts}
\usepackage{amssymb}
\newcommand{\text}{\mbox}
\def\openone{\leavevmode\hbox{\small1\kern-3.8pt\normalsize1}}
\def\RR{{\rm I\kern-.2emR}}
\def\tr{{\rm tr}\; }

\def\cs{{\cal S}}

\def\cf{{\cal F}}
\def\ci{{\cal I}}
\def\ca{{\cal A}}

\def\cz{{\cal Z}}

\def\cea{{\cal E}{\cal A}}

\newcommand{\ket}[1]{| #1 \rangle}

\newcommand{\beq}{\begin{equation}}
\newcommand{\eeq}{\end{equation}}
\newcommand{\beqa}{\begin{eqnarray}}
\newcommand{\eeqa}{\end{eqnarray}}

\begin{document}

\begin{center}
{\Large\bf Quantum information processing and quantum logic:\\
towards mutual illumination}\\
\bigskip
{\normalsize Howard Barnum}\\
\bigskip
{\small\it CCS-3, Mail Stop B256 \\
Los Alamos National Laboratories, Los Alamos, NM 87545
USA}
\\ {\tt email: barnum@lanl.gov} \\
\vskip 4mm
{\today}
\end{center}
\begin{abstract}
Quantum information and computation may serve as a 
source of useful axioms and ideas for the quantum structures 
project of characterizing and classifying types of physical
theories, including quantum mechanics and classical mechanics.  
The axiomatic
approach of quantum structures may help isolate what aspects of
quantum mechanics are responsible for what aspects of 
its greater-than-classical information processing power, and
whether more general physical theories may escape some common
limitations of classical and quantum theories.  Also, by 
by helping us understand how existing quantum algorithms work, 
quantum structures analyses may suggest new quantum protocols
exploiting general features of quantum mechanics.  
I stress the importance, for these matters, of 
understanding open and closed-system dynamics,
and the structure of composite systems in general
frameworks for operational theories, such as effect algebras,
convex sets, and related structures. 
\end{abstract}

\section{Introduction}

A main concern of the 
field of ``quantum logic'' and related algebraic structures has 
long been to understand the differences between quantum 
and classical physical theories by looking for axiomatic, 
often largely algebraic characterizations of
the two types of theories.  Usually, appealing 
axioms having an intuitively
understandable meaning have been sought.  One might hope to 
find a nice set $S$ 
of axioms which both kinds of theories satisfy, and
two other sets $Q$ and $C,$ 
such that $Q \cup S$ gives classical theory while $C \cup S$ gives
quantum theory.  Such a characterization of the two types of
theory would help one understand,
what they have in common and how they differ.  Changes to 
either the shared axiom set or the distinctive classical or quantum
axioms might help one understand what other kinds of 
physical theories are possible.  Such alternative kinds of theories
might be  
intermediate between the two types (perhaps keeping $S$ but
substituting some other set for $C$ or $Q$), 
or diverging from both on fundamental issues (perhaps differing on $S$).
I think it is fair to say that quantum logic and related fields have
made much progress towards this goal over the years, but that no 
characterization is yet widely recognized to have met it, although
some approaches have come close. 
Of course, the success of one approach would not mean
others should be abandoned:  it is possible, even likely, that 
there are various useful characterizations of quantum 
mechanics, classical mechanics,
and how they differ.

Over the last fifteen or twenty years,
the fields of quantum information theory (QIT) 
and quantum information
processing (QIP) have emerged. QIP involves physicists, computer scientists, 
mathematicians and others in investigating the fact that in a world
governed by quantum theory, one can do some tasks of {\em classical}
information 
processing faster or better than in a world governed by classical 
theory.  But in investigating these matters, one quickly comes to 
appreciate the importance of quantum {\em analogues}
of classical
tasks such as data compression and the transmission of states through
a noisy channel, in which the data or messages are themselves conceived
of as quantum in nature--as {\em quantum} information.
This has given rise to QIT, 
a quantum {\em generalization} of information theory, 
which is an analogue of the classical theory in which not just the
performance of classical tasks (possibly in superior fashion) by quantum
means is considered, but also the performance of tasks which are themselves
{\em quantum} and may or may not have an analogue in the classical world.

The prime examples of classical information processing by quantum means
are quantum computation and quantum 
cryptography.  Quantum computation allows
some tasks, such as factoring integers \cite{Shor94a}, \cite{Shor97a}, 
to be done in polynomial
time whereas no polynomial time classical algorithm is known despite
intense effort to find one.   It also allows brute-force search
(for instance, for solutions to problems in NP) to be done in 
the square root of the classical time 
\cite{Grover96a,Grover97a}, although
this cannot of course speed up an exponential search to a polynomial
one.  And quantum cryptography allows (with the aid of an authenticated
classical channel or a small amount of classical shared key to use in 
authenticating classical communication) the distribution 
of a large amount of key material  
in such a way that eavesdropping is virtually certain to be 
detected \cite{Bennett84a}---
a classically impossible task.  (The amount of key distributed can be
much larger than the amount required
for the classical authentication in the protocol.)

This has raised the question:  what aspects of 
quantum mechanics are responsible for these better-than-classical
feats of information processing?
Here there is scope for fruitful interaction between 
the fields of quantum information and quantum structures.  On the one 
hand, the axiomatic approach of QS
suggests
varying the axiom set and investigating the 
information-processing power of the resulting theories.  On the 
other hand, information theory and information-processing considerations
may suggest new and physically meaningful axioms,
for use in the quantum structures characterization program.

In this paper, I will investigate some of the most promising possibilities
for this interaction, and in some cases sketch how I think it might 
play out.  I will also, though tangentially, 
address the possible usefulness and limitations of 
such an approach as a way of gaining perspective on possible new physics,
for instance quantum gravity.

\section{An operational approach}
An important part of QS is concerned with what
I will call ``operational structures.''  This approach
goes back at least to the work of Mackey 
\cite{Mackey63a}.  I would include in it the work of
Ludwig and his school, the convex approach with important
contributions from
Ludwig, Davies, Edwards, 
Araki, Gudder, Pulmannov\'a, 
Beltrametti, and many others, and the ``test space'' and 
``effect algebras'' approach of Foulis and his coworkers and students,
Gudder, Schroeck, and (using initially different terminology) 
the Italian schools of Giuntini et. al., dalla Chiara,
and the Slovak school of 
Dvurecenskij, Pulmannov\'a and others.  (I also think the ``Geneva
school'' may have evolved in this direction, and there are category-theoretic
approaches which may also be related.)  I will use the term ``operational
structures'' for structures describing an abstraction of a physical
theory in terms of operations or measurements one may perform on a 
system, and the probabilities for outcomes of
these procedures.

In this paper I will sometimes adopt a version of the 
operational point of view which guides much of my own thinking 
in the area.  I will call it the ``procedural''
point of view since ``operational'' is already used in various
senses in quantum structures.  On this view, which I associate particularly
with Ludwig and his followers, but which has roots in the work of
Mackey or even earlier, 
the ``measurements'' which may ultimately be given some
formal structure such as that of effect algebra, test space, etc..,
are to be viewed as suitable equivalence classes of operational
procedures.  Any operation which we can perform on 
a system, including any sequence of two operations in a row, any 
interaction with an environment followed by an operation or measurement
on environment, system, or both, etc., 
should be included in the set of 
procedures from which equivalence classes are generated to give us an
effect algebra (or whatever structure we get).  I am investigating
a framework in which one begins with experimental procedures having
different possible outcomes, and divides out of the set of outcomes
a relation of ``probabilistic equivalence'' (two outcomes are equivalent
if they have the same probability in all physically preparable states).
One dubs the equivalence classes thus obtained ``effects'', 
because---as I will show elsewhere---one can
embed them into an effect algebra with natural relations 
(induced by the quotient map) to 
the Boolean algebras generated by the experimental outcomes themselves.
This effect algebra may in general contain 
additional effects which are not 
equivalence classes of the original outcomes.  If these are not 
to be countenanced, one obtains a structure I call a ``weak effect
algebra'':  all axioms of an effect algebra hold, except that 
the associative law is weakened so as not to require the existence
and equality of both sides of $(a \oplus b) \oplus c =
a \oplus (b \oplus c)$ whenever one side exists;  rather, equality
is to hold whenever both sides are defined.  
This notion of probabilistic equivalence has been criticized,
(e.g. in 
\cite{Cooke81a}, where it is attributed
to Bohr, and 
\cite{Wright78a}) 
but I believe it is nevertheless useful.
One complaint has been that 
outcomes which are equivalent in this respect may lead to different
probabilities for subsequent measurements \cite{Wright78a}.  
This is correct, but 
in the framework I am proposing, just means that
many different conditional dynamics can correspond to a given effect.
A complete theory, including dynamics and conditional dynamics,
will need structure beyond that of effects, 
but effect algebras are nevertheless an
important conceptual structure, probably adequate for statics
and providing a good starting point for the introduction of 
dynamical structure.  For example, see 
\cite{Foulis2000a} for a vigorous start on the project of introducing
at least reversible dynamics on effect algebras (via their ambient
unigroups), the last three 
sections of \cite{Wilce2000a} for similar considerations on test
spaces, and \cite{Pulmannova98a} for explicit discussion of conditional
dynamics in effect algebras with some convex structure.   
The construction of effects I have mentioned may well have been carried out
previously, by Ludwig, as part of a convex spaces approach to deriving
quantum mechanics.  I arrived at it as an abstraction of the lovely
presentation by Ludwig's student, Kraus, 
of the quantum case\cite{Kraus83a}.  

It is natural to suppose that
apparatus involving ``dice'' may be used to prepare (as a mathematical
idealization) any convex combination of states, and also to perform
any convex combination of measurement procedures.
 Even though for some
physical systems we may not have much freedom to actually perform
different preparations, dice or no dice, the supposition may apply
to many laboratory systems, and we might argue from some sort of
homogeneity considerations that the restrictions, if any, on 
theory derived from considering convex combination in laboratory
systems reveal general structure which we should suppose distant
systems, more difficult to manipulate, also to exhibit.  
In an effect algebra, requiring
that measurements can be convexly combined induces a notion
of multiplication of an effect by nonnegative real number
(and thus also a notion of convex combination of effects); 
since the orthosum $\oplus$ in most nice effect algebras is a 
restriction of the sum on a partially ordered 
Abelian group into which the effect algebra is embedded, 
one thereby gets a
real vector space structure for the effects.
\footnote{It is not 
completely clear to me yet how important convexity is for the
introduction of vector-space structure---Foulis and Bennett
\cite{Foulis94a} show the existence, for any field $K$
and effect algbebra $L$, of 
a ``universal vector space'' for $L$ over $K$.}
This
goes a long way towards enforcing a representation
of probabilities as linear functionals on a real vector space of 
effects, possibly a step on the way to a Born-like probability
rule representing probabilities as $\tr XT$ for linear operators
$X$ representing effects and $T$ representing states.  Obviously,
careful treatments must keep track of the ordering aspect of 
the relevant structures as well, and an appropriate representation 
theorem is proved in \cite{Gudder98b}.

The general operational approach {\em may} turn out to be a limiting, or 
limited, way of viewing physical theories.  It seems rather 
suited, however, to quantum mechanics, particularly ina Copenhagen
stylee, and
also adequate to treat classical mechanics in its
information-processing aspect.  The limitations arise because
this sort of theory takes ``operations,'' such as making a measurement
on a system, as basic terms in formulating a physical theory.  
It views a 
physical theory as a description of the behavior of a ``system,''
a part of the world viewed as susceptible to the performance of 
operations on it by, presumably, another ``part of the world,''
the observer or experimenter (\cite{Foulis98a}, p. ).  It seems, 
inherently, to renounce
the attempt to view the entire world as one structure, and physical
theory as describing that structure\footnote{In an interesting
article (brought to my attention by Chris Fuchs)
which relates it to the interpretation of quantum mechanics
Bilodeau\cite{Bilodeau96a} has termed
a similar distinction one between ``dynamical'' and ``geometric''
theories.}.  Inasmuch as many have suggested
quantum theory shows that project to be hubris, this is perhaps 
desirable.  Moreover, one might in some 
theories be able to view the entire world, if desired, as a system
of the kind the theory describes, with the ``procedures,'' or 
at least those viewed as measurements, imagined
performed by some external observer.  Classical theory may be such
a case.  Many believe that  quantum mechanics is not.  
However, Everett's ``relative-state'' interpretation
(the inspiration for what is sometimes called 
the ``many-worlds interpretation'') attempts 
such an ``objective'' interpretation of quantum mechanics.
Some, notably
Rovelli and Smolin \cite{LSmolin95a},
have proposed making the impossibility of such 
an external view an axiom for an acceptable physical theory.

Aside from quantum gravity aspects, the main interest the
Rovelli-Smolin
approach holds for me is, 
ironically, that it suggests a framework in which 
quantum mechanics is good for describing things from the point 
of view of subsystems, but not appropriate for the entire universe, 
but in which {\em nevertheless} there exists a mathematical 
structure---something like a topological quantum 
field theory (TQFT)---in 
which these local subsytem points of view are coordinated
into an overall mathematical structure which, while its terms may be
radically different from those we are used to, may still be viewed
as in some sense ``objective.''  However, it is far from clear yet
that this can be done while avoiding the more grotesque aspects
(proliferating macroscopic superpositions viewed as objectively
existing) and remaining conceptual issues (how to identify a preferred
tensor factorization, and/or preferred bases, in which to identify
``relative states'') of the Everett interpretation.

\section{Statics versus dynamics, composite systems, and the ``local'' 
nature of operational theories}
A large part of quantum structures has concentrated on the 
statics of states and measurements or propositions.  It is 
certainly reasonable to examine statics on its own, all the 
more so because in some cases the concern is with theories which 
may become part of a generalization of quantum theory, a model,
say for quantum gravity or a field-theoretic approach, rather
than a simple dynamical theory in which time is an external 
parameter.  In such a situation, it may be desirable to develop
a static algebraic structure first, and then to integrate it
into a dynamical, spacetime, or some more general physical 
framework.  The $C^*$-algebraic approach to quantum field 
theory \cite{Haag92a} \cite{Haag64a} \cite{Keyl97a} is one example of 
this, in which the local structures are taken to be $C^*$-algebras,
and they are to be coordinated into ``nets'' consistent with relativistic
spacetime structure.  The external-time Hilbert space dynamics 
of nonrelativistic quantum mechanics is 
another, more straightforward one.  Rovelli-Smolin 
is an attempt to coordinate ``local'' algebraic structure
(Hilbert space) into larger structures exhibiting some properties
of gravitation and quantum field theory.  However, it could 
also turn out (in my view this is very likely)
that the nature of the appropriate local structure requires input
from the global structure or the dynamics.  For one thing, we should
arguably be able to take an ``external'' view of the 
operations performed by one part of the global structure on another,
as dynamical interactions between two systems considered as a single
composite system, say.  It could even turn out that 
the composite of A and B is
not described by quite the same type of local theory as are A and B
themselves.  In fact, although in TQFT's the local theory type
is the same (complex inner product spaces), only in special
cases does composition of systems induce the tensor product of
the local systems.

For information-processing or computation, 
both dynamical considerations and composite systems
are of the utmost importance.  Since the environment which induces
noise in a system
or the apparatus used by an
information-processing agent must be considered together with the 
system,
a notion of composite system  is
needed.  And notions of composition are basic to
computational complexity, where the question may
be how many bits or qubits are needed, as a function of the 
size of an instance of a problem (number of bits needed to write
down an integer to be factored, say) to solve that instance.  Indeed,
the very notion of Turing computability is based on a factorization of
the computer's state space (as a Cartesian 
product of bits, or of some 
higher-arity systems), in terms of which a ``locality'' constraint 
can be imposed.  The constraint is, roughly, that only a few of these
subsystems can interact
in one ``time-step.''  The analogous quantum constraint 
allows only a few qubits to interact at a time.  In general operational
models, some notion of composition of systems, such as a tensor product,
together with a theory describing what dynamics can be  implemented
on a subsystem, could allow for circuit or 
Turing-machine models involving ``bits'' or other local systems of 
a nature more general than quantum or classical systems. 
There may be much to learn from a study of computational complexity
in such general systems.
For example, I think we are likely to find intuitively meaningful,
very general properties of operational physical theories, shared by
quantum and classical mechanics but also by a wider class of theories,
 which 
forbid, for conceptually clear reasons, exponential speed-ups of
brute-force search.
These 
properties may be linked to other properties of theories:  for example,
the second law of thermodynamics (impossibility of a {\em perpetuum
mobile}), or the impossibility of instantaneous signalling between
subsystems of a composite system.  
Richard Jozsa has suggested that the impossibility of such 
speedups could serve as a constraint on proposed 
new physics. 
In this regard, Lloyd and Abrams'\cite{Abrams98a} demonstration that 
theories of ``nonlinear quantum mechanics'' do permit such 
exponential speedups is relevant.  It is particularly interesting
in light of the fact that nonlinear quantum mechanics also appears
inconsistent with the second law of thermodynamics \cite{Peres89a}.
Moreover, Polchinski \cite{Polchinski91a} has 
argued that a class of nonlinear modifications of
quantum mechanics (which includes Weinberg's \cite{Weinberg89a} proposal)
either allow superluminal signalling (because
they allow instantaneous signalling between subsystems, which may
be spacelike-separated) or, if they do not allow superluminal 
signalling, allow something he calls an ``Everett phone.''  
This latter phenomenon involves a sequence of spin measurements,
in which the outcome of a later spin-measurement, conditional on
the first spin-measurement having resulted in spin up, depends
on what the observer {\em would have} proceeded to measure if
the first measurement had resulted in spin down. 
Polchinski discusses the Everett phone
in the framework
of the ``relative states'' interpretation of quantum mechanics.
It would appear to be inconsistent with a more standard view of 
quantum mechanics, for example as an ``operational theory'' of
the sort we have been considering.  On such a view, the probability
of a measurement result is taken to be independent of the context
in which it is measured---i.e., of the other outcomes tested.
(Such context-independence is automatic on the procedural point of
view described above: it is enforced by the definition of 
effects as probabilistic equivalence classes of measurement outcomes.)

The ``modern'' quantum formalism of 
``Positive Operator Valued Measures''---resolutions of unity, 
into positive operators $F_i \ge 0,~~\sum_i F_i = I$---
and completely positive maps on the space of operators on a Hilbert
space as representing the dynamics of a system conditional on getting
measurement result $i$, is the stock in trade of the quantum information
processing theorist.  The
CP-maps represent the most general physical
process that an encoder, decoder, or programmer initially independent
of the system can hope to get a system
to undergo.  CP-maps
may also represent the effect of an interaction with an environment on 
a quantum system.     Frequently, one looks for the optimal CP-map with 
respect to some information-processing criterion, or shows that no 
CP-map with certain information-processing properties exists.

There are several equivalent characterizations of CP-maps.  Write
$B(H)$ for the set of positive operators on a (finite-dimensional) 
Hilbert space $H$.  The definition of complete positivity
says that the map not only takes positive operators to positive operators, 
but that any extension of the map by the identity map on an additional 
system preserves the positivity of operators on the extended system.
That is, a CP-map $\ca$ satisifies $(\rho^{AB} \ge 0) 
\rightarrow \ci^B \otimes \ca^A (\rho^{AB})
\ge 0$.  If such a map is trace preserving, it represents a general 
open-system dynamics not conditioned on any measurement outcome or additional
information about the $B$ system.  If it is trace-decreasing, it may represent
dynamics conditional on a measurement outcome (the trace of the 
output operator
when a trace-one operator is put in, is the measurement probability).
An equivalent characterization of TPCP
maps is that the map's action on 
$B(A)$ is obtained by considering a {\em reversible} dynamics, i.e.
a unitary operation, on $A$ and some {\em environment} $E$, but considering
its effect only on $A$.  An equivalent characterization for trace-decreasing
maps is as the effect, on $B(A)$, of unitary interaction between $A$ and $E$
(so that elements of $B(A)$ are conjugated by $U^{AE}$)
followed by the action of an operator $G^E \otimes I^A$
(also by conjugation), representing a measurement made on the 
environment.  (This corresponds to an effect $G^\dagger G$ on the
environment, and the trace-decreasing condition imposes 
$G^\dagger G \le I$.   It is easily seen
that $G^\dagger G$ is all that matters
to the operation realized on the system.)  For arbitrary CP-maps (with no 
trace constraints), $G$ may be an arbitrary operator on $E$.)  
Still another characterization of $CP$-maps, this one referring only to the 
system $A$,  but having little {\em prima facie} intuitive motivation,
is that their action may be 
written $\ca(X) = \sum_i A_i X A_i^\dagger$.  Trace-preservation is
the condition $\sum_i A_i^\dagger A_i = I$.

A condition for exploring information-processing in 
operational theories
is an understanding of the dynamics we may use to achieve our information
processing goals (and that nature, or an adversary, may use to foil us).  
A promising approach is to generalize one or more of the above 
characterizations of $CP$-maps.  The first two,
which are the ones whose conceptual significance is clearest, are the
easiest to generalize.  Doing so
requires notions of composite
system and of dynamics on such systems.  Additionally,
the first requires a notion of reversible dynamics, 
while the second requires
the notion of extension of subsystem dynamics by the identity.
The third characterization of CP-map, though not requiring a notion
of composite system,  probably requires more algebraic
structure on a single system than do the first two, since it makes use
of the product of operators.  

A notable example of the introduction
of additional structure is Gudder and Greechie's introduction of the
{\em sequential product} on effect algebras, which abstracts some
properties of the product $A * B = A^{1/2} B A^{1/2}$ on effect 
algebras.  We have
$\tr A^{1/2} B A^{1/2} \rho = \tr B^{1/2} A^{1/2} \rho 
A^{1/2} B^{1/2}$, which is the probability of measuring the effect
$B$ after
obtaining the effect $A$ from a prior measurement, when
the conditional dynamics of the first measurement are
the ``square-root dynamics' $\rho \mapsto A^{1/2} \rho A^{1/2}$.
The square-root dynamics is a natural generalization of the
projection postulate and {\em ``L\"uder's rule''} to effects that
are not projections, and it shares important features with 
them.  For instance, when a given resolution of unity is measured,
the square-root dynamics are the conditional 
dynamics that cause the least expected 
disturbance to an initial pure state picked uniformly from Hilbert
space \cite{Barnum98d}, \cite{Barnum2001a}.  

\section{Composite systems, and tensor products}
As mentioned above, I believe that the ``procedural'' approach
to operational theories leads naturally to theories whose statics
is described by an 
effect algebra. This notion is simultaneously an abstraction of the unit
interval of operators $0 \le e \le I$ on a (say, finite-dimensional) 
Hilbert space, and the 
space of positive valued functions $f$ bounded above by unity
on a (for convenience, let us say finite) set.  (The latter represents
possibly ``fuzzy'' classical measurement outcomes.)  An effect
algebra is a set on which is defined a partial binary operation 
$\oplus$, sometimes called the orthosum.  (In effect algebras of 
Hilbert-space operators, $\oplus$ is just addition of operators, 
except that it is not defined when $X + Y > I$, since then $X +Y,$
while positive
is not in the unit interval.
The operation is commutative and associative in the strong
sense that when the effect referrred to by 
one side of the commutative or associative law is
defined, so are the effects mentioned on the other side, 
and they are equal.  
Additionally the algebra contains special elements
$0$ and $1$.  These satisfy $x \oplus 0 = x$, and $x \oplus y = x 
\rightarrow y = 0$.  Importantly, for every effect $x$ there is exactly one
other effect, its {\em orthosupplement} $x^\perp$, which can be added
to it to get $I$. 

The category of effect algbebras, $\cea$, 
presents aspects which are quite fascinating and appealing, 
but some of those same
aspects also suggest that we may need to focus on some
subcategory
of $\cea$, especially
in order to focus on differences between quantum and classical 
information processing.  A consideration of composite systems
allows us to define relatively
natural notions of tensor product of effect algebras\footnote{This 
may be done using the standard category-theoretic definition
of tensor product in terms of universal properties of bimorphisms;
for QIT theorists unfamiliar with this construction, \cite{Geroch85a}
Chapters 1 and 7, are an excellent introduction}.  For effect
algebras, the is due to Dvurecenskij \cite{Dvurecenskij95a}; for
a nice alternative construction of this tensor product see 
Wilce \cite{Wilce94a}\cite{Wilce98a}.  These constructions
are similar to 
Foulis and Randall's \cite{Foulis81a}, \cite{Bennett84a},
\cite{Randall81a} 
pioneering 
construction of the tensor product of (most) test spaces and
Foulis and Bennett's construction \cite{MKBennett93a} 
for orthoalgebras.  This 
notion 
may be derived (in one of several ways) by essentially requiring 
that it include all elements of the cartesian product of the
two factors, that the partial operation ``$\oplus$'' of the
algebra be such that all locally implementable measurements
(consisting in doing a measurement on one system and, conditional
on the result of that measurement, doing a measurement on the other
system) are measurements in the product algebra.  All states on
the new effect algebra will automatically add up to one on such 
measurements.  
Remarkably, the tensor product may also 
be derived from the requirement that
it contain as a subset the cartesian product $E \times F$,
exhibit no instantaneous influence in either direction, 
and that it be a universal object satisfying these requirements

Beautiful as this construction is, when we apply it to Hilbert space
effect algebras, we see that it does not give the effect algebra
of the tensor product Hilbert space.  For the algebra of sharp
Hilbert-space effects (projectors), this was noted by  \cite{Randall81a},
\cite{Wilce90a}, but it clearly extends to the algebra of all effects.
There are states 
on this tensor product reproducing the statistics of all the quantum
states, including the entangled ones, as far as tensor product measurements
are concerned.  But there exist other states, given by tracing
with Hermitian operators that are positive on pure tensors
but negative on some entangled outcomes, 
with statistics different from
the quantum ones even for product measurements.
``Entangled measurements'' (perhaps we could abstract
this notion!) are not available in this structure, so negativity
on entangled outcomes does not prevent these from being
legitimate states on the tensor product of the effect algebras.
An independent construction of what is essentially the 
Dvurecenskij tensor product  
of two Hilbert space effect algebras 
appears in \cite{Fuchs2001b}, \cite{Fuchs2002a}.  

It would be fascinating
to investigate the computational 
power of a model starting with, say, Hilbert effect algebras of a
given dimensionality and building up composite systems via the 
DW tensor product.  The competition between the availability of
more states and the availability of fewer measurements makes this
particularly interesting.  
But it also suggests the need for a new kind of operational structure---a 
new category, perhaps---that will give us the correct notion of tensor
product in the quantum-mechanical case as well as the classical case.
A variety of additional requirements have been imposed on effect
algebras---convex structure, ``S-domination,'' existence of a sequential 
product---and these seem to get us closer to the Hilbert space structure
(while still leaving open the possibility of classical effect algebras).
But probably 
none is quite satisfactory yet, either for 
for characterizing the Hilbert space effect algebras, or for 
the somewhat different project I have suggested here, of creating
a category (call it little category $\cz$) with the property (call it VOOM!)
that the subcategory of 
Hilbert-space effect algebras is closed under formation of the $\cz$ tensor
product.  With collaborators, I am attempting to find such a
$\cz$ which is a subcategory 
of the category of effect algebras.  

In either $\cea$ or $\cz$,  we could proceed to investigate general 
open-system conditional dynamics.  On the other hand,
dynamics could provide a source of requirements helping to define the
category $\cz$.  For instance, we could require that the first two definitions
of complete positivity, when appropriately 
generalized to the category $\cz$, coincide.
We could investigate the {\em generalized measurement problem} in $\cz$.
For starters, we could require that any measurement $M$
that can be done on 
$A \in \cz$ can also be done by 
interacting $A$ (probably reversibly) with $B$ and measuring $B$
(in the sense that both the measurement probabilities and the conditional
dynamics associated with $M$ 
can be achieved thus), and vice versa, that any procedure of the
type: reversibly interact and then measure the apparatus $B$ corresponds
to a measurement in $\cz$.

Sequential measurements raise 
the question of the relation of the operations considered to 
the parameter of time.  Up to now, I have considered operations independently
of the length of time they take;  and that is perhaps reasonable, since 
the time they take may depend on how they are implemented.  But clearly 
this is an important question for complexity theory.  A related question
is whether the set of operations implementable at one time is the same 
as that implementable later.  Perhaps this is the operational version of the
question of time
translation invariance.  In both classical and quantum theory 
the set of operations is time-translation invariant.
In a general setting, this property is likely 
to make  for a more tractable and probably more
interesting theory.
Possibly we will want to get by, in a 
truly operational theory, without an absolute notion of time:  rather, 
we might define time by using some subsemigroup of dynamical operations
as a clock (cf. e.g. \cite{Rovelli90a}).

\section{Distinguishability}

An important tool in quantum information theory, and QIP theory, 
has been measures of distinguishability of two, possibly mixed,
quantum states.  A copious supply of such measures may be obtained
in a general operational setting using a strategy which has 
proved useful in quantum information theory.  We starts
by considering classical measures of distinguishability of probability
distributions.  
Say our operational theory
consists of a subcategory $\cf$ of the category of effect algebras,
together with a convex set $\cs$ of states on algebras in 
this subcategory, and a set of possible dynamical
evolutions on this subcategory (probably a convex 
subset of the morphisms of the subcategory).  
We simply define the distinguishability
of two states $\rho, \omega \in \cs$ as the maximum over
effect-tests $\Sigma$ (sets of effects $e_i$ such that
$\oplus_i e_i = 1$) of the 
distance between the classical probability distributions
$p_{\Sigma,\rho}$ and $p_{\Sigma,\omega}$ 
induced by $\rho$ and $\omega$
on the outcomes in $\Sigma$:
\beqa
D_{\cf}(\omega, \rho) := \max_{\Sigma} 
D_{cl}(p_{\Sigma, \rho},p_{\Sigma, \omega})\;.
\eeqa

The question then arises:  
when $D_{cl}$ is nonincreasing
under classical dynamics,  
is the induced $D_{\cf}$ also 
nonincreasing, under the notion of dynamical evolution 
incorporated into $\cf$?
I suspect a counterexample can be found even in the quantum
case.  There has been extensive study of the distance measures
which are contractive under the morphisms appropriate to 
the operational category of standard nonrelativistic quantum
systems (i.e., unital completely positive linear maps 
on operators interpreted as observables, or,
dually, trace-preserving completely positive linear maps
on the state space of such systems).  Petz \cite{Petz96a}
and Lesniewski and Ruskai \cite{Lesniewski99a} are particularly noteworthy. 
It may well be that the contractiveness of an appropriate set
of distance measures is a principle that (combined with 
a tensor product structure of system composition which may well,
as it does in the cases of orthoalgebras and effect algebras,
automatically prohibit instantaneous inter-system influence)
renders exponential speedup of brute-force search impossible. 

Other concepts applicable to classical probability, 
such as entropy, can similarly be extended to the noncommutative
case.  Entropy may prove a bit tricky because, in the quantum-mechanical
case, it needs to 
be minimized over finegrained, sharp  measurements
(rank-one projectors), not just all measurements.
Thus one might need to 
define sharp and finegrained measurements in $\cea$;
\cite{Gudder98a} and \cite{DallaChiara95a} considered these notions;
also, the notions of {\em principal} and {\em characteristic} elements
(cf. \cite{Foulis2000a}) may be relevant.

\section{Information-disturbance tradeoffs and quantum cryptography}
One of the most remarked-on differences between quantum and classical
mechanics is that in quantum mechanics 
our gaining information about the identity of an 
unknown state in general disturbs it.  This difference is
between the quantum {\em analogue} of a classical task, 
and the classical task.  (Should one encode classical information into
quantum states so that it can be reliably read out, 
as different states of an orthonormal basis, for instance,
the information about which of these states obtains {\em can}
be extracted without disturbing them.)  Closely related is the
impossibility of {\em cloning} an unknown quantum state. (If you
could make an independent copy of an unknown quantum state, you could
measure it and extract information without disturbing the original.)
Quantitative expressions of this information-disturbance relationship
as a monotonic tradeoff
have been developed, for example, in \cite{Fuchs95b},
\cite{Barnum98d}, \cite{Barnum2001a} and \cite{Banaszek2001a}.  
Such tradeoffs could be studied
in more general settings.  The information-disturbance
tradeoff lies at the heart of the possibility of eavesdropping-proof 
quantum distribution of classical secret key.  
Such protocols use
nonorthogonal states (or similar devices involving entanglement) to 
encode prospective key-material.  
An eavesdropper who gains information about the key will reveal 
herself by the disturbance she causes to these states (some of
which Alice and Bob sacrifice, by measuring them and discussing
the results in order to detect disturbance).  After they are
satisfied there has been no eavesdropping, Alice and Bob are
able to use the states, via some measurements and some discussion,
to generate shared, secret, classical key.

At one time, some thought that quantum mechanics could be used for 
another purpose that is known to be classically impossible without
relying on assumed computational limitations of the parties:
bit committment.  In a bit commitment protocol, Alice can perform
some action which assures Bob that Alice has committed to a choice
of value (0 or 1) for a bit.  At some later time, Alice can perform
an action which reveals that bit to Bob. In a successful protocol, 
nothing Bob can do allows him to gain more than a negligible
amount of information about the bit before Alice reveals it, and nothing
Alice can do allows her more than a negligible probability of being
able to change the bit after she has committed to it.   
This is impossible in quantum mechanics, as in 
classical mechanics \cite{Mayers96a}, \cite{Lo97a}.  
The impossibility of bit-committment
is another candidate for the axiom-set $S$, 
of information-processing limitations shared by quantum 
and classical mechanics.  This was suggested by Gilles 
Brassard;  Chris
Fuchs and Brassard  
also speculated that the combination of no-bit-committment 
and eavesdropping-proof key distribution might single out quantum
mechanics.  This intriguing suggestion is spun into a thought-provoking
fantasy in \cite{Fuchs2001a}.

\section{Axioms revisited}

In the axiomatic approach to characterizing classical and quantum
theories
mechanics, one meets several types of axioms.  Some, usually
in the shared set $S$,  set up a general
framework for operational theories:  
axioms for effect algebras, convexity, and the like.
Others might be dubbed ``conceptual,''.  One task 
conceptual axioms frequently do is distinguish classical from 
quantum mechanics.  These are axioms like the existence,
in some abstract sense, of superpositions of states
or the
requirement, within the convex approach,
that the state space be a Choquet simplex (that the 
decomposition of states into extremal states be unique, which holds
classically but not quantum-mechanically).
Many axioms abstracted from quantum information 
considerations will fall into this class---axioms 
such as the existence of information-disturbance tradeoffs for 
ensembles of states with certain properties, or axioms about the
behavior of measures of distinguishability of states.  
The ``conceptual''
content of these axioms varies from extremely operational
and task-related (possibility/impossibility of specific types of 
protocols) through an intermediate level (conceptual properties 
viewed as probably responsible for certain tasks, e.g. information-
disturbance tradeoffs) to relatively abstract and mathematical
notions (such as the Choquet simplex axiom, or perhaps no-cloning) 
which nevertheless naturally arise 
when one looks at states abstractly as information.

It is likely, however, that in addition to such axioms
a characterization
of quantum and classical mechanics will involve what I will call 
``regularity axioms.''  Probably these form a continuum with 
``conceptual'' axioms, rather than a clear dichotomy, but the polarity
is still relevant.  Regularity axioms may 
impose a high degree of symmetry on structures, possibly through
the introduction of a group action on them
\cite{Wilce2000a}, \cite{Foulis98a}.  Another example of regularity 
appears 
in Hardy's derivation of quantum mechanics. Here, t
he convex structures approach
(redeveloped independently in a simple 
finite-dimensional version) gets one, as it has often 
been used to do, representations of probabilities as linear functionals.
Then two key notions of ``dimensionality'' $N$ 
of a system (defined in an operational way not immediately dependent on 
the vector space notion, though of course in the end related to it)
and of ``number of degrees of freedom'' $K$ are introduced.
$K$ represents the number
of measurement outcomes whose probabilities one needs to know in order
to know the entire state (the probabilities of {\em all} measurement
outcomes).  In quantum mechanics, $K$ 
is the square of the dimension:
one needs the probabilities of a set of effects which are linearly 
independent in the real vector space of 
Hermitian operators.  Hardy postulates that a theory 
(or theory-type) must have a ``universal'' relation between $K$
and $N$, specifying one as a function of the other.  I consider this
an example of a ``regularity'' axiom, and it goes a very long way
towards pinning us down to a choice between classical and quantum
mechanics.  Hardy's composite system axiom, that $K$ and $N$
for composite systems are the
products of those for the subsystems, has the flavor of regularity,
but it may also arise conceptually, through as part of the notion of
composite system---perhaps through an intimate relation of this
notion with the category-theoretic one of tensor product.
This is the sort of thing that deserves more investigation.  In fact,
one may sometimes discover alternative formulations of a regularity
axiom, or of its conjunction with other axioms, that give it a more
conceptual interpretation.  The prospect of such discoveries is yet 
another reason for pursuing 
the axiomatic approach to operational theories.  

I suspect that even after much conceptual content has been wrung
from axioms initially motivated by regularity or symmetry, regularity
and symmetry axioms will still remain an essential part of axiomatic
formulations of quantum and classical mechanics as well as  generalizations 
and alternatives.  Investigation of the ``irregular'' structures that
arise when they are relaxed will be an important part of understanding
their meaning, and the meaning of the conceptual axioms that impose
less regularity.  For example, while finite-dimensional quantum and
finite classical systems both obey Hardy's universal $K-N$ relationship,
the category of quantum systems with superselection sectors does not.
(Of course, in other senses this is still a fairly ``regular'' category
of theories, and I am not aware of a completely conceptual characterization
of it.)  

\section{Conclusion}
I hope to have convinced both the 
quantum structures community and the more abstractly inclined 
sectors of the quantum computation/information
community that there are important contributions to both fields
to be made by collaboration and exchange of tools, methods, and ideas.
A crisper delineation of what makes certain
quantum protocols work, using notions from quantum logic, may
not only lead to clearer understanding of these protocols
but may
suggest new ones.  And considerations
from quantum information, whether they involve 
the abstract approach of
considering various kinds of operational theory as embodying different kinds
of ``information,''  or the more 
concrete approach of considering the power of
different kinds of 
operational theories to do particular tasks of classical 
information-processing, provide 
ideas for use in the quantum structures project of characterizing, 
classifying, and inventing types of operational
physical theories.  Critical to this project will be understanding
the interplay between composition of physical systems, and the
dynamics of those systems, in order to generalize 
the ``modern'' quantum mechanical formalism of effect-valued measures
and completely positive maps, so effective in quantum information
theory, to more general operational settings.

\section*{Acknowledgements}
Discussions with Carlton Caves, Dave Foulis, 
Chris Fuchs, Lucien Hardy, Richard Jozsa, and Alex Wilce, 
among others, have influenced
my thoughts on these matters.  I thank the IQSA and
the organizers of the 2001 IQSA
meeting in Cesena for the opportunity to travel to and speak at the
meeting.  I thank the US Department of Energy and the European Union 
(under the
QAIP consortium, IST-1999-11234) for funding while the work resulting
in this paper was being done.


\bibliographystyle{IEEE}



\end{document}